\documentclass[11pt,a4paper]{article}
\usepackage{graphicx}
\usepackage{jheppub}
\usepackage{setspace}
\usepackage{amsmath,amssymb,color,mathrsfs,verbatim,bbm,wasysym,pstricks,epsfig,slashed}

\newcommand{\cO}{\mathcal{O}}
\newcommand{\cL}{\mathcal{L}}
\newcommand{\Z}{\mathbb{Z}}
\newcommand{\im}{\mathrm{Im}}
\newcommand{\N}{\mathcal{N}}
\newcommand{\tr}{\mathrm{Tr}}
\newcommand{\R}{\mathbb{R}}
\newcommand{\F}{\mathbb{F}}

\title{Resurgent Supersymmetry and String Theory}
\author[a,b,c]{Christopher Brust}
\affiliation[a]{Department of Physics and Astronomy, Johns Hopkins University, \\
Baltimore, MD 21218}
\affiliation[b]{Department of Physics, University of Maryland, \\
College Park, MD 20742}
\affiliation[c]{Perimeter Institute for Theoretical Physics\\
31 Caroline Street North, ON N2L 2Y5, Canada}
\emailAdd{cbrust@perimeterinstitute.ca}
\abstract{We study a realization of accidental supersymmetry in type IIB string theory as a proof-of-principle of the mechanism and as a prototype of the strong sector present in resurgent supersymmetry, a warped UV-completion of natural supersymmetry. We first introduce the mechanism of accidental supersymmetry as a way of producing a supersymmetric spectrum in the IR of a quasi-conformal field theory, and then go on to discuss the utility of the mechanism in the specific BSM model of resurgent supersymmetry. The realization of accidental SUSY that we study is IIB string theory on an orbifold of the Klebanov-Strassler solution.}
\arxivnumber{1408.4833}

\begin{document}

\begin{flushright}UMD-PP-013-010\end{flushright}

\maketitle
\flushbottom


\section{Introduction}
\label{sec:intro}

The strongly coupled regimes of gauge theories are the home of many diverse phenomena in quantum field theory which are often missed in perturbative studies of those theories, such as confinement and the growth of extra dimensions. Strong coupling has been used as a tool for realizing several mechanisms in field theory, and such lines of inquiry have suggested new solutions to the hierarchy problem \cite{Randall:1999ee, Contino:2003ve, Agashe:2004rs}. However, with the recent discovery of a SM-like Higgs boson \cite{Chatrchyan:2013lba, Aad:2013xqa} and the absence of any discoveries of new physics to stabilize the electroweak scale, this motivates even more strongly the need to test the postulate of naturalness very thoroughly. Recently, there have been many models which add in a minimal module of new physics capable of stabilizing the ``little hierarchy'' from the electroweak scale up to a new physics scale above the reach of the LHC. The supersymmetric version of this resolution to this ``little hierarchy problem'' has been dubbed ``more minimal'' or ``natural'' SUSY \cite{Dimopoulos:1995mi, Cohen:1996vb, Brust:2011tb, Papucci:2011wy}, and LHC search strategies capable of probing such a scenario have been studied extensively \cite{Brust:2012uf, Cao:2012rz, Evans:2012bf, Bi:2012jv, Han:2012cu, Franceschini:2012za, Bai:2013ema, Duggan:2013yna, Bai:2013xla, Kim:2014yaa}.

However, with such a low cutoff for such models, one very rapidly requires a UV completion in order to study the viability of the model. Supersymmetric extensions are certainly possible, but can themselves come with UV tuning issues \cite{Arvanitaki:2013yja}, motivating one to consider alternative UV completions. It seems only natural to consider UV completions of natural SUSY which themselves involve the ingredients of those mechanisms which solve the big hierarchy problem. Strong coupling is such an ingredient, and indeed, using it to solve the little hierarchy problem has been proposed \cite{Sundrum:2009gv} and studied in the context of the Randall-Sundrum (RS) framework \cite{chrisraman} through the use of the AdS/CFT correspondence \cite{Maldacena:1997re}. We will refer to this framework as {\it resurgent supersymmetry}.

Resurgent supersymmetry builds off of the idea that the Higgs boson could be a light composite of a nonsupersymmetric quasi-conformal field theory \cite{Strassler:2003ht}. In a general QFT with a supersymmetric matter content but no supersymmetry, the flowing of scalar masses to be small can sometimes lead to a supersymmetric theory in the IR, despite the fact that the flow there was nonsupersymmetric. We will refer to the mechanism at work in such a construction as {\it accidental supersymmetry}, defined generally as a feature of quantum field theories in which supersymmetry is an accidental symmetry of the IR. Although accidental SUSY itself is a broad-reaching subject with many possible uses, what we concern ourselves with primarily is its utility in BSM physics in the model of resurgent SUSY. The appreciation of the value of accidental SUSY in BSM model-building goes back to works including \cite{Gherghetta:2003he, Goh:2003yr}. It was further argued in \cite{Sundrum:2009gv} that the use of an accidentally supersymmetric strong sector could be utilized in BSM model building in order to keep a select few superpartners light in order to reproduce the bottom-up minimal natural SUSY framework \cite{Brust:2011tb}. 

However, by virtue of being a model in RS, the story of resurgent supersymmetry is necessarily not UV-complete due to the presence of higher-dimensional operators in the 5d Lagrangian in order to reproduce the proper low-energy, 4d physics, and therefore the physics responsible for the UV completion must appear imminently at higher energies. This is in fact true of almost every model in a warped 5d setting, and so therefore it is paramount to ascertain whether UV completions of {\it any} mechanism in RS {\it even exist}, resurgent supersymmetry being no exception. Beyond the comfort of having a UV completion in hand, studying even a part of the space of UV completions might offer insight into generic non-minimal low-energy phenomenology. In this paper, we describe details of the construction of an accidentally supersymmetric strongly coupled sector in IIB string theory, and offer a specific string-theoretic model which serves as a prototype of the strongly coupled sector in resurgent SUSY.

The string model we study in this paper is a $\Z_2$ orbifold of the Klebanov-Strassler solution. It satisfies several criteria necessary to be labelled as a prototype of the strong sector present in resurgent SUSY. Specifically, it is approximately accidentally supersymmetric; there are no relevant operators in this orbifold. Consequently, we can consider a family of 4d effective theories below some UV scale $\Lambda$ where we nonsupersymmetrically deform the Lagrangian of the supersymmetric orbifold theory in a parametrically small fashion by perturbations which respect both the continuous and discrete symmetries of the theory. Because none of these deformations are relevant, this SUSY-breaking does not ``transmit'' well into the IR, a notion to be made more precise in the following section. Furthermore, the Klebanov-Strassler solution as well as its orbifold which we study in this work smoothly confine at some small scale, breaking chiral symmetry. Below this scale, in the deformed orbifold theory, we have an approximately supersymmetric spectrum of composites.

Note that it makes no difference in this model {\it how} the Lagrangian deformations came to be at the scale $\Lambda$. The virtue of this construction is that we can be assured that regardless of what the UV SUSY-breaking dynamics (above the scale $\Lambda$) are, {\it all} such deformations which respect the symmetries give rise to an accidentally supersymmetric spectrum. Furthermore, as we argue momentarily, the model has various phenomenological deficiencies, including a lack of dynamical 4d gravity. If we were to fix these deficiencies, then doing so may have a hand in some of the details of SUSY-breaking, negating the relevance of a discussion of such details here. However, the crucial point is that regardless of how this model is UV-completed, all of its UV completions {\it necessarily} exhibit accidental SUSY.

Our model is not without its deficiencies. Due to the difficulty level in obtaining precisely the MSSM as a low-energy theory in string theory, we do not attempt to reproduce precise properties of the MSSM (including the gauge group and matter sector) in this work, although if advances were to be made in string model building, more phenomenologically accurate models of accidental SUSY would be an interesting direction to pursue. Furthermore, the strong sector we exhibit is not coupled to 4d gravity, owing to the noncompactness of the conifold in the KS solution. As a first step, this is not a problem, though it is an issue that should be remedied in future more phenomenologically accurate models. In order to maintain a warped throat but compactify the space, adding 4d gravity, we would be forced into working either with F-theory or its orientifold limit. We do not study an F-theoretic completion here, though it would be interesting to study in future work.

The outline of the paper is as follows: in section \ref{sec:accsusy} of this work, we will construct the mechanism of accidental supersymmetry in a completely general setting, and then in section \ref{sec:resusy}, move on to discuss details of an accidentally supersymmetric sector suitable for the story of resurgent supersymmetry. Readers concerned with the details of field-theoretic model building may opt to read through section \ref{sec:accsusy}, whereas readers interested in BSM physics and its appearance in string theory may elect to instead begin in section \ref{sec:resusy}.

In section \ref{sec:asst}, we shift our attention from details of accidental and resurgent SUSY in field theory to details of the AdS-dual string theory. We discuss necessary ingredients for the construction of accidentally supersymmetric models in IIB string theory, culminating in a proof-of-principle example of the mechanism; as stated earlier, the model is IIB string theory on a $\Z_2$ orbifold of the Klebanov-Strassler solution, also described as a warped product of $\R^{1,3}$ with a deformed complex cone over the zeroth Hirzebruch surface $\F_0$. We outline the necessary computations in section \ref{sec:f0}, and expound upon relevant details as well as some background material in the appendix.

Because this model serves as a prototype of the strongly coupled sector in resurgent supersymmetry, this model also serves as evidence meriting further study of the mechanism of accidental SUSY in the context of RS. We do not attempt to make our stringy model fully realistic because of the great technical challenge in realizing precisely the MSSM in string theory (see, e.g., \cite{Malyshev:2007zz} and references therein), but merely attempt to identify features of our model which can act as proxies for a fully realistic construction. Although we choose to focus on type IIB string theory as our UV-complete framework, it would be interesting to pursue realizations of accidental SUSY in other UV-complete frameworks in the future.


\section{Accidental SUSY}
\label{sec:accsusy}

In this section, we describe the mechanism of accidental supersymmetry in generality, before moving on to discuss its role in the story of resurgent supersymmetry. Before proceeding, however, we clarify some notation. We work in a $d$-dimensional quasi-CFT, viewed as an effective field theory valid below some scale. A scalar primary operator in the quasi-CFT which is a singlet under all global symmetry groups (including $U(1)$s and discrete symmetries) is known as a {\it global singlet operator} (GSO). If the operator has scaling dimension $\Delta < d$, it is said to be a global singlet {\it relevant} (primary) operator (GSRO), following the notations of \cite{Strassler:2003ht}. Likewise, an operator with scaling dimension $\Delta = d$ is a global singlet {\it marginal} (primary) operator (GSMO).

Throughout this work, in order for us to maintain computational control of the theory, we need to be in a large CFT-color $N$ limit, or more generally, a regime in which conformal perturbation theory is valid.

We will be classifying primary operators to leading order in $\frac{1}{N}$ (or more generally, the AdS perturbation expansion), but use the terms GSRO and GSMO to refer to the leading-order scaling dimensions. If the $O(\frac{1}{N})$ corrections to scaling dimensions make a marginal operator relevant or irrelevant, we say that the operator is {\it $\frac{1}{N}$-marginally relevant} or {\it $\frac{1}{N}$-marginally irrelevant}\footnote{These definitions should not be confused with the standard definitions of marginally relevant or irrelevant, though our definitions are conceptually similar to those.}.

{\it Accidental supersymmetry} describes field theories in which supersymmetry is an accidental symmetry in the IR of an RG flow. Accidental SUSY can come in a number of different varieties; it can be exact or approximate, and weak or strong. We now expound upon this terminology in more detail.

\begin{figure}[ht]
\centering
\includegraphics[width=10cm]{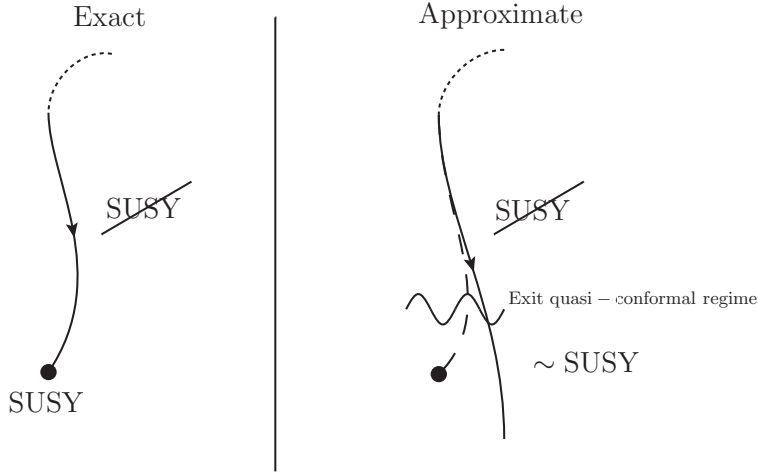}
\caption{\label{fig:rgflows}Schematic IR RG flows associated with strong exact (left) and approximate (right) accidental supersymmetry. We do not impose any requirements on the UV. In the exact flow, the flow itself is nonsupersymmetric (despite having a supersymmetric matter content), but ends on a supersymmetric fixed point. In the approximate flow, we flow towards a supersymmetric fixed point, but exit the quasi-conformal regime before reaching the fixed point, leaving an approximately supersymmetric field theory of composites below $\Lambda_{comp}$.}
\end{figure}

{\it Exact strong accidental supersymmetry} is a feature possessed by quantum field theories which have a nonsupersymmetric flow that ends on a supersymmetric fixed point\footnote{Of course, there is nothing wrong with ending on a fixed surface of dimension $> 0$ instead.}, as illustrated on the left in figure \ref{fig:rgflows}. The flow into the UV of models of strong accidental SUSY could be either supersymmetric or nonsupersymmetric, as described in \cite{Sundrum:2009gv}. If SUSY is broken in the UV in a way that respects the global symmetries of the theory, then at some high scale $\Lambda$, but still below the scale of SUSY-breaking, one can parameterize the effects of SUSY-breaking by adding SUSY-breaking spurions with arbitrary coefficients $\varepsilon_i$ multiplying every GSO\footnote{Note that we mean to study every scalar operator, not every scalar superoperator; we do not want these GSOs to be added in a way that respects supersymmetry. Although there is nothing wrong with doing so, it is not accidental SUSY.} to the Lagrangian:

\begin{equation}\cL(\Lambda) = \cL_{IR~SCFT} + \sum_{\mathrm{GSOs}~i}\varepsilon_i(\Lambda) \cO_i\end{equation}

\noindent As we run into the IR down to a scale $\mu$, the Wilson coefficients run:

\begin{equation}\cL(\mu) = \cL_{SCFT} + \sum_{\mathrm{GSOs}~i}\varepsilon_i(\Lambda) \left(\frac{\mu}{\Lambda}\right)^{\Delta_i-d} \cO_i + \ldots \end{equation}

\noindent where the ellipses refer to higher-order corrections in $\varepsilon$. We see that if {\it all}\footnote{It is important to note that if not all $\Delta_i$ are larger than $d$, one could find flows to supersymmetric fixed points by tuning those $\varepsilon_i$ to $0$. This is not what we mean by accidental SUSY; we are interested in {\it natural} models, where the fixed point is attractive from all directions.} $\Delta_i > d$, then as we continue flowing into the IR, SUSY is parametrically restored since our SUSY-breaking terms flow to $0$. Therefore, to say that a model possesses strong accidental SUSY is equivalent to saying that the QFT does not possess GSROs or GSMOs.

In weakly coupled QFTs, anomalous dimensions of operators are perturbatively small, and so if there are any scalars $\phi_a$ in the theory (possibly transforming nontrivially under global symmetry groups), then $|\phi|^2$ is always a GSRO. Therefore, in general, one needs to be strongly coupled to eliminate GSROs, as anomalous dimensions can be $O(1)$. However, strong coupling alone is not enough; one requires strong coupling all along the flow in order for SUSY-breaking effects to flow to zero. Therefore, most simply, the theory must sit near a conformal fixed point in order to exhibit accidental SUSY.

Despite the beauty of strong accidental SUSY, the requirement that the flow end on a strongly-coupled fixed point is not helpful for BSM model building, where there exists a mass gap. However, it is possible to have {\it approximate strong accidental SUSY} instead; such models would only be approximately conformal for some duration of IR flow, and at some dynamically generated, hierarchically small ``compositeness scale'' $\Lambda_{comp} \ll \Lambda$, the theory leaves the quasi-conformal regime. This could occur when {\it supersymmetry-preserving} but {\it non-conformal} operators $\cO^{SUSY}_j$ are generated in the low-energy Lagrangian:

\begin{equation}\cL(\Lambda_{comp}) = \cL_{IR~SCFT} + \sum_{SUSY~j} \lambda_j(\Lambda_{comp}) \Lambda_{comp}^{d-\Delta_j}\cO^{SUSY}_j + \sum_{\mathrm{GSOs}~i}\varepsilon_i(\Lambda) \left(\frac{\Lambda_{comp}}{\Lambda}\right)^{\Delta_i-d} \cO_i + \ldots\end{equation}

\noindent where $\lambda_j$ are dimensionless small parameters representing deviations from conformality. Without SUSY-breaking in the UV, the theory would have looked approximately superconformal at high energies, then would have had a gap at $\mu \sim \Lambda_{comp}$, with a weakly-coupled description given in terms of SQCD below $\Lambda_{comp}$. However, once we include SUSY-breaking terms, the low-energy theory depends on whether there exist and whether the theory has generated SUSY-breaking GSROs or GSMOs. If there had been no GSROs or GSMOs, then we would have been suppressing SUSY-breaking effects along the flow, and below the compositeness scale, the resulting field theory could be described by an approximately accidentally supersymmetric theory of light ``composites''.

{\it Exact} and {\it approximate weak accidental supersymmetry} are defined in a similar fashion to strong accidental SUSY, but only insisting that there are no GSROs, allowing the possibility of GSMOs. In such models, any SUSY-breaking that was present in the UV could still be present in the IR, but could remain parametrically small, controlled by a perturbatively small parameter $\varepsilon$ (such as the SUSY-breaking scale divided by the messenger scale, possibly dressed by loop factors) sitting in front of a GSMO in the Lagrangian. Such models are useful for BSM model building as they provide a way of transmitting a small amount of SUSY-breaking to the IR.

It is possible that a model possesses weak accidental SUSY in the planar limit, but that $O(\frac{1}{N})$ corrections spoil this effect, resulting in $\frac{1}{N}$-marginally relevant GSOs. In this case, we write $\Delta = d + \gamma$ where the anomalous dimension\footnote{Normally in field theory, the anomalous dimension refers to twice the difference between the true scaling dimension and the engineering dimension of an operator. In this context, we mean a slightly different definition; we mean the difference between the true scaling dimension and the {\it leading order in $\frac{1}{N}$}-scaling dimension, or in AdS, the size of the quantum correction to the field's tree-level mass.} $\gamma$ is $O(\frac{1}{N})$ and negative. In such cases, one must worry that the effects of the operator remain perturbative as we flow into the IR. If we parameterize the UV coupling $\varepsilon$ as $\lambda \Lambda^{d-\Delta}$, where $\lambda$ is dimensionless and may be parametrically small, then the IR coupling will be given by

\begin{equation}\cL \supset \lambda \Lambda^{-\gamma} \left(\frac{\mu}{\Lambda}\right)^{\gamma} \cO \end{equation}

\noindent At scale $\mu$, this matches onto a description in terms of the effective IR coupling $\varepsilon' = \lambda' \mu^{d-\Delta}$:

\begin{equation}\cL \supset \lambda' \mu^{-\gamma} \cO\end{equation}

\noindent We would like this IR coupling to remain perturbative all the way down to $\mu = \Lambda_{comp}$; i.e. $\lambda' \ll 1$. Performing the matching, we see that this is equivalent to

\begin{equation}\lambda' = \lambda \left(\frac{\Lambda_{comp}}{\Lambda}\right)^{2\gamma} \ll 1 \qquad \mathrm{or} \qquad \Lambda_{comp} \gg   \lambda^{-\frac{1}{2\gamma}} \Lambda\end{equation}

\noindent In $\frac{1}{N}$-marginally relevant realizations of resurgent SUSY, this will constrain how large a hierarchy there can be between the SUSY-breaking scale $\Lambda$ and the cutoff scale $\Lambda_{comp}$ of the (B)SM EFT.

We emphasize that is a fairly generic concern in such models, because such models often come with global symmetries with associated conserved currents $j_\mu$ with protected scaling dimension 3. These currents have scalar superpartners\footnote{Just as the gauge boson $A_\mu^a$ is sourced by some current $j_{\mu}^a$, so too is its scalar superpartner, the auxiliary $D$-term, sourced by some scalar current $j_s$.} $j_s$ in the IR SQFT with protected scaling dimension 2. In the case of a $U(1)$ global symmetry, this scalar superpartner could be a deadly GSRO, but could be eliminated with a discrete symmetry. However, if the global symmetry is nonabelian, then $j_s^a$ carries some global symmetry index $a$ and is not a GSRO. Nevertheless, it can still be worrisome, because it can be used to build a double-trace GSMO at the planar level $j_s^a j_s^a$, which could possibly receive negative $O(\frac{1}{N})$-corrections to its scaling dimension, making it $\frac{1}{N}$-marginally relevant.

\section{Resurgent SUSY}
\label{sec:resusy}

We now discuss the story of resurgent supersymmetry. We first outline the requirements of the model, then expound upon these points below. The framework has three sectors:

\begin{itemize}
\item The MSSM minus the third-generation quark and Higgs superfields, which we call the ``elementary'' sector
\item An approximately superconformal strongly coupled sector exhibiting weak approximate accidental SUSY subject to the constraints listed below, and has composites which serve as third-generation quarks and Higgses
\item A SUSY-breaking sector which breaks SUSY somewhere well above $10$ TeV
\end{itemize}

\noindent The accidentally supersymmetric sector should satisfy the following checklist:

\begin{itemize}
\item A large-$N$ expansion
\item It should not contain GSROs
\item It should contain global symmetries that can be weakly gauged to serve as the SM gauge groups
\item It should not be destabilized by the weakly-coupled (MSSM-like) sector, as studied and described in \cite{Sundrum:2009gv}
\item It should exit the quasi-conformal regime at $\Lambda_{comp} \sim 10$ TeV
\end{itemize}

We would like to couple the MSSM to an approximately superconformal sector in such a way that the third-generation quark and Higgs sectors are composites. The MSSM gauge groups are global symmetries of the approximately superconformal sector, which are then weakly gauged\footnote{Note that our previous discussions of accidental SUSY focused on the spectrum of the approximate SCFT; accidental SUSY is even broader-reaching because a ``focusing effect'' can push us towards an accidentally supersymmetric spectrum in the weakly gauged external sector as well \cite{Sundrum:2009gv, chrisraman}.}. We couple the SUSY-breaking sector to the elementary sector directly. The SM fermions remain light because of chiral symmetry, but the elementary sfermions obtain large masses. Finally, as in \cite{Sundrum:2009gv}, we arrange for the gauginos to also be light by model-building an R-symmetry to forbid gaugino mass-terms. The R-symmetry is to be broken at a scale $\lesssim \Lambda_{comp}$, giving the gauginos masses.

The SUSY-breaking sector also communicates with the approximately superconformal sector, but the approximately superconformal sector possesses weak approximate accidental supersymmetry, and so below the compositeness scale, the quasi-conformal sector (containing the Higgs, Higgsinos, and third-generation quarks and squarks) is approximately supersymmetric. In doing so, we have arranged for the spectrum to resemble that of natural SUSY \cite{Brust:2011tb} in the IR.

\begin{figure}[ht]
\centering
\includegraphics[width=7cm]{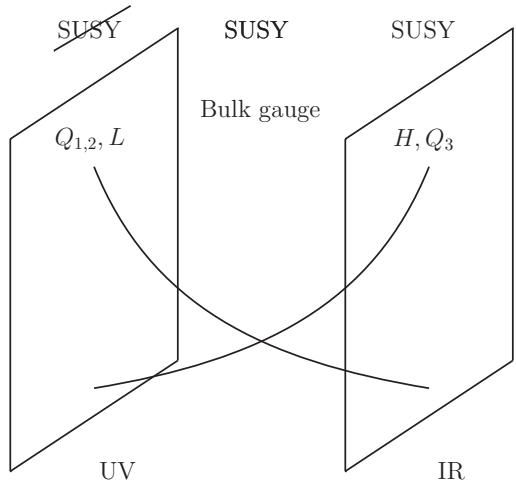}
\caption{\label{fig:rs}The RS description of the BSM model proposed in section \ref{sec:accsusy}.}
\end{figure}

The above discussion has been in terms of four-dimensional physics; through the use of the AdS/CFT correspondence, we can also discuss higher-dimensional dual descriptions of the above framework. Such a description is naturally best formulated in the language of RS. The dual description of the above model is portrayed in figure \ref{fig:rs}. We break SUSY on the UV brane, but preserve it in the bulk and on the IR brane. The Higgs and third-generation quark superfields are IR-localized, whereas all other matter fields are UV-localized. Therefore, the low-energy theory contains light third-generation squarks and Higgses. The gauge superfields propagate in the bulk. Models of this sort have been explored, e.g. in \cite{Gherghetta:2003he, Gherghetta:2011wc}.

For completeness, we repeat the checklist of criteria our model should satisfy to be accidentally supersymmetric, but on the gravity side:
\begin{itemize}
\item A small-coupling expansion
\item The supergravity should not contain tachyons capable of transmitting SUSY-breaking, and the radion should be stabilized with the Goldberger-Wise mechanism or something comparable
\item The RS bulk should contain gauge fields
\item RS loop effects\footnote{Note that it is not just one-loop or two-loop effects which are important; verifying stability requires the inclusion of all-loop diagrams, a feat which can be accomplished with the holographic RG (see, e.g., \cite{Sundrum:2009gv}).} should drive weakly gauged sectors towards accidental SUSY as well
\item There should be an IR brane at a position corresponding to an energy scale of $10$ TeV
\end{itemize}

Finally, for dynamical 4d gravity, we should have a UV brane, on which we break SUSY.


\section{Searching for Accidental SUSY in String Theory}
\label{sec:asst}

In this section, we outline requirements for building a prototype of an accidentally supersymmetric strong sector by restricting its dual string theory, viewed as the UV-completion of the RS model described above. In section \ref{sec:f0}, we go on to find such a prototype; it is an orbifold of the Klebanov-Strassler string theory.

The goal of this paper is to provide a realization of an approximately superconformal sector which exhibits approximate weak accidental supersymmetry. Although such a model can be obtained readily in RS, it requires a UV-completion due to the presence of higher-dimensional operators in the Lagrangian, and so we pursue a realization of such a sector in IIB string theory instead. Furthermore, the technology that has been developed to study string theory and its supergravity approximation provide computational control over the model at hand, but in a fashion which assures us that a full UV-complete description of the physics exists. We now discuss how to convert our gauge-description requirements into string-description requirements.

The goal will be to have a gauge-description model which preserves SUSY, and then break SUSY by hand at high energies and have it return as we flow to the compositeness scale. Such a model requires a strongly coupled quasi-SCFT, which is a warped throat with a quasi-AdS geometry in the dual string description. In order for the low-energy theory to possess dynamical four-dimensional gravity, the dimension which is warped should not be noncompact like the AdS geometry, but rather a one of six compactified dimensions comprising a manifold $M_6$ (while nevertheless having a warped region). In order for supersymmetry to be preserved in the bulk, $M_6$ should be a Calabi-Yau threefold. We would like to utilize the supergravity approximation to the full string theory, and so we work at small string coupling. This ensures that SUGRA modes remain light (and are therefore dual to operators with small scaling dimensions) but make the string modes heavy, and therefore not dual to GSROs/GSMOs. The existence of a mass gap (making our accidental SUSY approximate) is dual to the finiteness of the throat. Finally, Lagrangian deformations of the SCFT are dual to classical linearized perturbations to a supergravity solution, and the scaling dimensions and representations of the primary operators we deform the Lagrangian by are determined by the radial scaling and representations of the perturbations of fields in supergravity.

However, right away we run into trouble, because there is an no-go theorem \cite{Giddings:2001yu} stating that the only way to obtain a warped throat in type IIB string theory on a {\it compact} $M_6$ is through the presence of O3-planes or D7-branes. In the absence of local objects in the theory, the only solution to the equations of motion are vanishing $G_3$ and constant fiveform flux and warp factor, giving an unwarped solution. Consequently, in order to have a warped solution in string theory, and therefore a quasi-CFT in the dual gauge theory, one needs to add local objects such as O3-planes or D7-branes which violate the assumptions of the no-go theorem. Therefore, one would hope that one could find warped models of accidental SUSY in F-theory compactifications \cite{Vafa:1996xn}, or in the orientifold limit of F-theory \cite{Sen:1997gv}.

However, these constraints do not apply to warped, noncompact geometries. Although these models are perfectly satisfactory in their own right, they lack dynamical four-dimensional gravity, and therefore are unsatisfactory for attempts to recreate complete BSM physics models. Nevertheless, these geometries can act as noncompact ``linearizations''\footnote{More precisely, noncompact geometries which geometrically behave as good approximations to the compact geometry in some region of spacetime.} of full compact solutions. The bulk of these compact manifolds act as ``UV branes'' on the EFTs living in the throat. Therefore, one can study the physics of a throat alone, and incorporate fluctuations sourced by physics in the bulk of the compact manifold in the effective description. This methodology is described and utilized in, e.g., \cite{Baumann:2009qx, Baumann:2010sx} to study the potentials of D3-branes for the purposes of studying inflation in string theory. It is advantageous to use this formalism because it frees us from needing to discuss the specifics of SUSY-breaking, allowing us to be sure that we've captured all possible methods of transmission of SUSY-breaking into the IR.

By carrying out a study of noncompact Calabi-Yaus as outlined above, we are not guaranteed the existence of a compact F-theory solution (giving rise to dynamical 4d gravity) which reproduces the result of using the noncompact geometry. However, if such a solution does exist, we are ensured that we have captured the relevant physics to the extent that the noncompact ``linearization'' is valid. Furthermore, F-theory compactifications such as the ones described in \cite{Giddings:2001yu, Kachru:2009kg} give hope that an appropriate compactification should exist to reproduce the model discussed below. Finally, the strong sector's global symmetry groups in our model below, which can be weakly gauged to serve as prototypes of the SM gauge groups, are based on isometries of the noncompact Calabi-Yau. However, when we compactify F-theory on a Calabi-Yau fourfold, there cannot possibly be isometry groups to weakly gauge, because in general, compact Calabi-Yaus do not possess continuous isometry groups. Therefore, we require symmetry groups from another source, such as brane constructions, in our F-theory construction. These issues are beyond the scope of this work, but must be considered when offering an F-theory description of accidental and resurgent SUSY, and so we offer additional discussion in section \ref{sec:conclusions}.

We are led, therefore, to consider a warped space which is of the form

\begin{equation}ds^2 = e^{2A(r)} \eta_{\mu\nu} dx^\mu dx^\nu + e^{-2A(r)}ds_{M_6}^2 \end{equation}

\noindent with

\begin{equation}ds_{M_6}^2 = dr^2 + r^2 ds_{X_5}^2 \end{equation}

\noindent where $ds_{X_5}^2$ is the metric on a five-dimensional horizon manifold. In order for the noncompact space $dr^2 + r^2 ds_{X_5}^2$ to be Calabi-Yau (thereby preserving bulk supersymmetry), $X_5$ should be Sasaki-Einstein \cite{Morrison:1998cs}. As a full F-theory construction complete with SUSY-breaking is quite a challenging feat, we settle for the more modest goal of looking for noncompact solutions to supergravity with brane sources, bulk supersymmetry and with this metric ansatz which have no GSROs, deferring the question of realizations on compact manifolds until future work.

Supposing we had such a background to work with, classical perturbations of the various supergravity fields by non-normalizable solutions to the linearized equations of motion are dual to deformations of the Lagrangian of the dual SCFT. Of course, such solutions would be normalizable upon compactification of the manifold. These non-normalizable modes would include, for example, the elementary particles of the MSSM. The masses of the linear supergravity perturbations determine the scaling dimensions of the dual primary operators, and the quantum numbers of the global symmetry group follow as well. By the AdS/CFT correspondence, those objects which are dual to Lagrangian perturbations of the CFT are scalar fields on $AdS_5$. Therefore, one can classify all scalar single-trace primary operators, their scaling dimensions (to leading order in $\frac{1}{N}$) and their global quantum numbers by studying the UV perturbation theory of 5d scalars about the background in question. Note that these supergravity field perturbations will be sourced by SUSY-breaking sources in the compact part of the manifold and therefore do not need to respect supersymmetry.

There is a famous example of a noncompact, finite warped throat in string theory; this is the Klebanov-Strassler (KS) solution \cite{Klebanov:2000hb}, which describes IIB string theory on a warped deformed conifold. It is a perturbation away from the fixed point of the warped throat of Klebanov-Witten (KW) \cite{Klebanov:1998hh}. KW describes string theory on a warped product of $\R^{1,3}$ with the conifold, a real cone over the space $T^{1,1}$. KW describes a theory with $N$ D3-branes placed at the tip of the conifold, whose worldvolumes coincide with $\R^{1,3}$. The dual CFT is a strongly-coupled fixed point of the quiver diagram shown on the left in fig. \ref{fig:kwksquivers}. Discussions of $T^{1,1}$ and the KW solution are given in the appendix.

\begin{figure}[ht]
\centering
\begin{tabular}{cc}
\includegraphics[width=5cm]{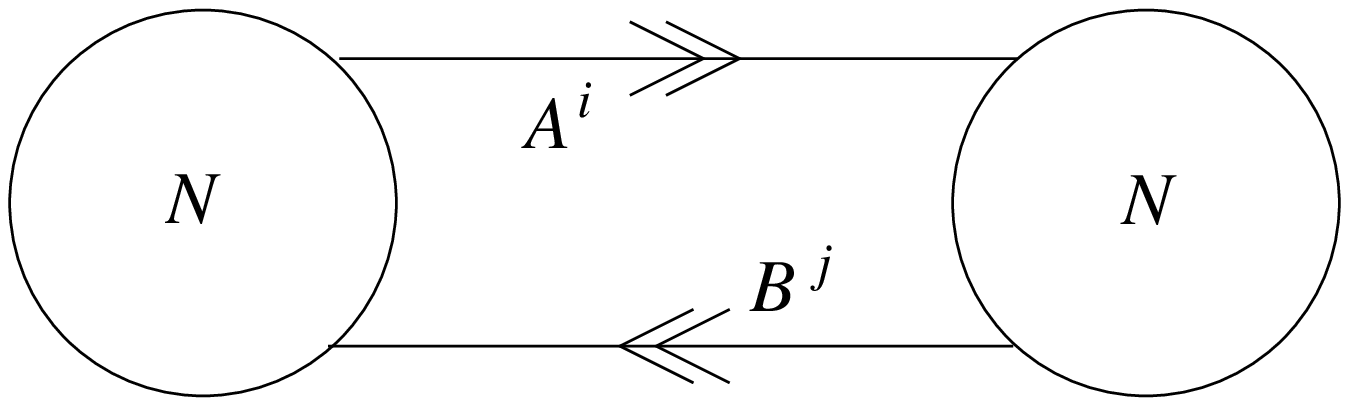}
&
\includegraphics[width=5cm]{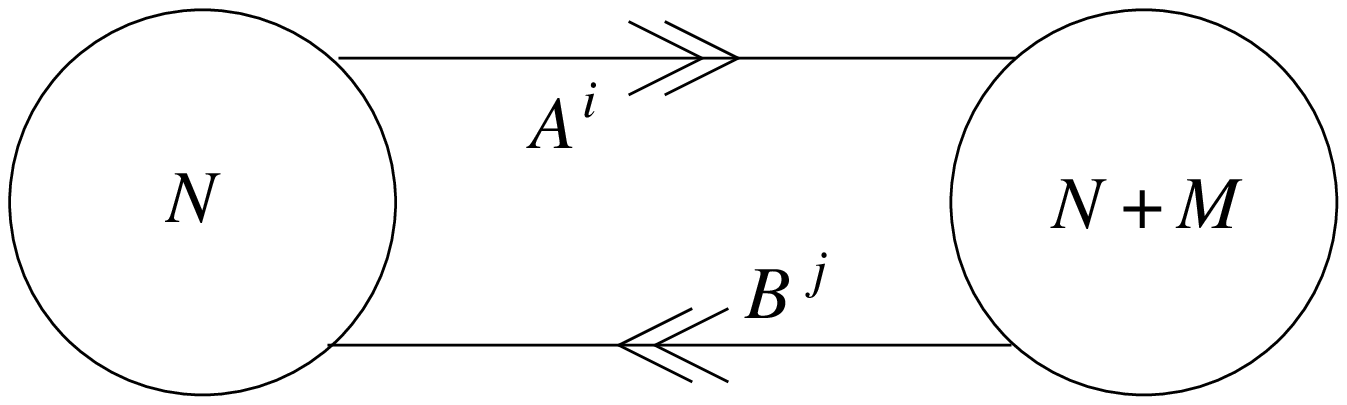}
\end{tabular}
\caption{\label{fig:kwksquivers}The quiver diagrams associated with the Klebanov-Witten (left) and Klebanov-Strassler (right) theories.}
\end{figure}

The KS solution is obtained by modifying the KW solution by wrapping $M$ D5-branes (with $M \ll N$) around the twocycle at the base of the conifold, and allowing the remaining three directions to be parallel to the D3-branes. This action famously deforms the conifold and generates a logarithmic dependence of the total flux through $T^{1,1}$ on the radial coordinate, which is dual to a gauge theory with gauge groups $SU(N) \times SU(N+M)$. Due to nonvanishing beta functions, this gauge theory cascades as we flow into the IR, undergoing repeated Seiberg dualities and returning to self-similar states, but with lower-rank gauge groups, finally ending deep in the IR when we've exhausted all possible Seiberg dualities. The finiteness of the throat is AdS/CFT dual to dimensional transmutation in the gauge theory, and is what will be responsible for the ``approximate'' nature of accidental SUSY in our model discussed below. 

KS contains a warped throat with a nontrivial supersymmetric field content in the IR, and an ``IR brane'' that comes to be via dimensional transmutation, making it a promising starting point for a study of accidental SUSY in string theory. However, compactifying and breaking SUSY generically in the UV will lead to a violent disruption of IR SUSY, in particular due to the presence of the GSRO $|\tr(AB)|^2$ ($\Delta = 3$) in the spectrum of operators \cite{Kachru:2009kg}. Consequently, KS does not exhibit accidental SUSY. Therefore, we will be interested in an orbifold of KS that removes this operator from the spectrum, as under our orbifold, $\tr(AB)$ is odd.
 

\section{A $\Z_2$ orbifold of Klebanov-Strassler}

\label{sec:f0}
\begin{figure}[ht]
\centering
\includegraphics[width=6cm]{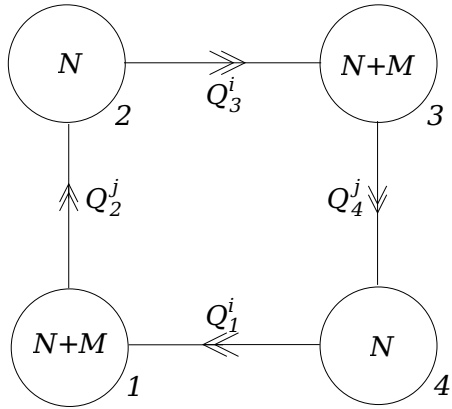}
\caption{\label{fig:orbifoldquiver}The quiver diagram associated with the theory dual to the complex cone over $\mathbb{F}_0$.}
\end{figure}

There is a close cousin of KW which describes string theory on a warped product of $\R^{1,3}$ with the complex cone\footnote{More precisely, we consider the noncompact total space of a line bundle fibered over $\mathbb{F}_0$, where the line bundle is the anticanonical bundle of $\mathbb{F}_0$.} over $\mathbb{F}_0$; this is just a $\Z_2$ orbifold of the conifold\footnote{This space can also be viewed as a real cone over $T^{1,1,2}$.} \cite{Morrison:1998cs}. The orbifold can be described on the supergravity side by taking the KW solution and modding out by $z_i \sim -z_i$. In terms of $T^{1,1}$, this operation can be described as identifying $\psi \sim \psi + 2\pi$. Note that this is nontrivial because the coordinate range of $\psi$ is $0 \leq \psi < 4\pi$ on the conifold. This is dual to orbifolding the gauge group and $A \sim -A$ in the gauge dual, removing the dangerous $\tr(AB)$ from the spectrum. The resulting theory has four gauge groups and four types of matter field, as illustrated by the quiver diagram in figure \ref{fig:orbifoldquiver}. In addition, there are some additional $U(1)$s which we will discuss more below. More details of the orbifold are given in the appendix.

When we carry out this orbifold, we insist that whatever F-theory construction we use respects two particular $\Z_2$ symmetries: first, a $\Z_2$ outer automorphism of KW (inherited by the orbifold theory) where we exchange $A$ and $B$ and exchange the two gauge groups (referred to simply as $\Z_2$ in table \ref{tab:matching}), and second, the $\Z_2$ in the orbifold theory obtained by setting the gauge couplings of gauge groups 1 and 3 equal as well as the gauge couplings of gauge groups 2 and 4 equal. We insist on the first so that the $U(1)_B$ scalar current Tr$(\bar{A}e^{V_1}Ae^{-V_2} - \bar{B}e^{V_2}Be^{-V_1})$ is not in the spectrum, and we insist on the second because it ensures that the theory is conformal, as well as enforcing that the deformed theory's RG flow is KS-like rather than chaotic.

This solution can also be deformed and exhibits KS-like running. In terms of the basis of oneforms described in the appendix, all $g_i$ are invariant under this operation, and therefore so are the fluxes in KS. This theory has been studied before, but in a different context \cite{Franco:2003ja, Franco:2004jz, Franco:2005fd}. Also note that this is to be contrasted with the SUSY-breaking orbifold described in \cite{Kachru:2009kg}, as our orbifold preserves SUSY. As discussed in section \ref{sec:asst}, the idea is to categorize all non-normalizable perturbations of the supergravity solution, and therefore gain an understanding of the scaling dimensions of possible Lagrangian deformations of the dual gauge theory. If one wanted, one could construct the nonsupersymmetric solutions where one explicitly broke SUSY by deforming by these GSOs, flowed into the IR and studied the effects of these deformations. However, by employing the framework discussed in section \ref{sec:accsusy}, we see that regardless of the details of the low-energy spectrum, it is approximately accidentally supersymmetric, which was our goal. Therefore, we can instead simply study the supersymmetric (undeformed) theory in order to demonstrate our point. Furthermore, we don't even need to study the deformed conifold; we can relate the computation in KS to a computation of the spectrum of primary operators in KW, as we now discuss.

Our only goal is to obtain the effective $AdS_5$ masses of all non-normalizable perturbations of the string theory on the deformed complex cone over $\mathbb{F}_0$ due to compactifications which behave as $AdS_5$ scalars, as these are the perturbations which will be dual to scalar perturbations of the dual gauge theory Lagrangian. Studying the perturbation theory around the deformed cone would in general be a true feat, but fortunately, since we're only interested in UV Lagrangian perturbations, dual to perturbations at moderate\footnote{We are working at large radii relative to the deformation parameter of KS, but still in the near-brane limit $r \ll R$.} $r$, we can utilize the fact that the geometry becomes approximately that of $AdS_5 \times T^{1,1}/\Z_2$, and study perturbations on that geometry instead. Furthermore, since the action of the orbifold becomes much more transparent on the gauge side, it is generally sufficient to classify perturbations in KW and study the action of the orbifold after AdS/CFT matching. This procedure has been carried out extensively before for KW/KS \cite{Ceresole:1999zs, Ceresole:1999ht, Baumann:2010sx}; we present the results below, and review some details of the derivation in the appendix. However, there are some subtleties related to additional $U(1)$s which should be treated with care, which we briefly elaborate on and discuss in more detail in the appendix.

In the original KW theory, there is a vector-like ``baryon number'' $U(1)$ symmetry where we assign $A$ charge $1$ and $B$ charge $-1$. This $U(1)$ is inherited by the orbifold theory, but due to the division of $A$ into $Q_1$ and $Q_3$ and $B$ into $Q_2$ and $Q_4$, additional $U(1)$s are present in the daughter theory. We denote these $U(1)_{B_A}$, where $Q_1$ has charge $1$ and $Q_3$ has charge $-1$, and $U(1)_{B_B}$, where $Q_2$ has charge $1$ and $Q_4$ has charge $-1$. These symmetries are in fact gauge symmetries that arise from the orbifold of $SU(2N)\rightarrow SU(N)\times SU(N)\times U(1)$. There are mixed anomalies of these $U(1)$s with the $SU(N)$ gauge groups, which are ``cancelled'' by the generalized Green-Schwarz mechanism, giving masses to the $U(1)$ gauge bosons. Even still, these $U(1)$s leave behind residual {\it global} symmetries which give rise to currents with scalar superpartners, possibly introducing GSROs not present in the parent theory. However, despite the fact that the Green-Schwarz mechanism is active, the global $U(1)$s {\it remain anomalous}, and so their associated currents have scaling dimensions which are not protected. Finally, in the dual string theory, the supergravity has no modes which could be dual to conserved currents, because the orbifold acts freely on $T^{1,1}$ and so does not give rise to new massless twisted-sector string modes. Therefore, we suspect that those currents of the $U(1)$s are in fact dual to massive twisted-sector string modes, and so are not GSROs. We discuss this issue further in the appendix.

{\renewcommand{\arraystretch}{1.1}
\begin{table}[h] \centering \small
\begin{tabular}{|c|c|c|c|c|c|c|}
\hline 
Scalar $\cO$ & Type & $\Z_2$ & Orb. & $\Delta$ & SUGRA & $(j,l,r)$ \\ \hline \hline
$I$ & - & E & Y & 0 & Vac. & $(0,0,0)$ \\ \hline
$\mathrm{Tr}(AB)$ & Chiral & E & N & $1.5$ & $\Phi_-$ & $(\frac{1}{2},\frac{1}{2},1)$ \\ \hline
Tr$(\bar{A}e^{V_1}Ae^{-V_2} - \bar{B}e^{V_2}Be^{-V_1})$ & Cons. & O & Y & 2 & $G_3^{II}$ & $(0,0,0)$ \\ \hline
Tr$(\bar{A}e^{V_1}Ae^{-V_2})$ & Cons. & - & Y & 2 & $\Phi_-$ & $(1,0,0)$ \\ \hline
Tr$(\bar{B}e^{V_2}Be^{-V_1})$ & Cons. & - & Y & 2 & $\Phi_-$ & $(0,1,0)$ \\ \hline
$\mathrm{Tr}(AB)_{\theta^2}$ & Chiral & E & N & 2.5 & $G_3^I$ & $(\frac{1}{2}, \frac{1}{2}, -1)$ \\ \hline
Tr$(\bar{A}e^{V_1}Ae^{-V_2} - \bar{B}e^{V_2}Be^{-V_1})_{\theta^2}$ & Cons. & O & Y & 3 & $G_3^{I}$ & $(0,0,-2)$ \\ \hline
Tr$(\bar{A}e^{V_1}Ae^{-V_2})_{\theta^2}$ & Cons. & - & Y & 3 & $G_3^I$ & $(1,0,-2)$ \\ \hline
Tr$(\bar{B}e^{V_2}Be^{-V_1})_{\theta^2}$ & Cons. & - & Y & 3 & $G_3^I$ & $(0,1,-2)$ \\ \hline
Tr$(ABAB)$ & Chiral & E & Y & 3 & $\Phi_-$ & $(1,1,2)$ \\ \hline
Tr$(W_1^2 + W_2^2)$ & Chiral & E & Y & 3 & $G_3^{III}$ & $(0,0,2)$ \\ \hline
Tr$(W_1^2 - W_2^2)$ & Chiral & O & Y & 3 & $h$ & $(0,0,2)$ \\ \hline
$\sim \tr(ABe^{-V_1}\bar{B}\bar{A}e^{V_1})$ & Long & E & Y & 3.29 & $\Phi_-$ & $(1,1,0)$ \\ \hline
Tr$(\bar{A}e^{V_1} A BAe^{-V_2})$ & Cons. & - & Y & 3.5 & $\Phi_-$ & $(\frac{3}{2}, \frac{1}{2}, 1)$ \\ \hline
Tr$(\bar{B} e^{V_2} B ABe^{-V_1})$ & Cons. & - & Y & 3.5 & $\Phi_-$ & $(\frac{1}{2}, \frac{3}{2}, 1)$ \\ \hline
Tr$(W_{1\alpha} AB + AW_{2\alpha} B)_\theta$ & Chiral & E & N & 3.5 & $G_3^{II}$ & $(\frac{1}{2}, \frac{1}{2}, 1)$ \\ \hline
Tr$(ABAB)_{\theta^2}$ & Chiral & E & Y & 4 & $G_3^I$ & $(1,1,0)$ \\ \hline
$\tr (W_{1\alpha} A e^{-V_2}\bar{A}e^{V_1} +W_{2\alpha} e^{-V_2}\bar{A}e^{V_1} A)_\theta$ & Semicons. & - & Y & 4 & $G_3^{II}$ & $(1,0,0)$ \\ \hline
$\tr(W_{1\alpha} e^{-V_1}\bar{B} e^{V_2}B+W_{2\alpha} B e^{-V_1}\bar{B} e^{V_2})_\theta$ & Semicons. & - & Y & 4 & $G_3^{II}$ & $(0,1,0)$ \\ \hline
$Leg\left(\frac{1}{g_1^2}+ \frac{1}{g_2^2}\right) =$ Tr$(W_1^2 + W_2^2)_{\theta^2}$ & Chiral & E & Y & 4 & $\tau$ & $(0,0,0)$ \\ \hline
$Leg\left(\frac{1}{g_1^2}- \frac{1}{g_2^2}\right)=$ Tr$(W_1^2 - W_2^2)_{\theta^2}$ & Chiral & O & Y & 4 & $A_2 = \omega_2$ & $(0,0,0)$ \\ \hline
\end{tabular}
\caption{A matching of SUGRA and SCFT scalar modes with $\Delta \leq 4$. The columns include: scalars in their respective superoperators, why their dimensions might be protected, whether they are even under the outer $\Z_2$-automorphism of KW, whether they are present in our orbifold theory, their scaling dimension to leading order in $\frac{1}{N}$, what mode they are dual to in SUGRA, and their quantum numbers under $SU(2)\times SU(2)\times U(1)_R$. $Leg$ refers to the Legendre-transform of an operator, as discussed in the appendix.}
\label{tab:matching}
\end{table}
}

We have elucidated the matching of all scalar SUGRA perturbations and scalar SCFT primaries with $\Delta \leq 4$ in table \ref{tab:matching}. In order for an operator to be dangerous to us, it must be a GSRO and furthermore be even under the orbifold action $A \rightarrow -A$, $B \rightarrow B$. Furthermore, it must be even under the particular outer automorphism $\Z_2$ symmetry discussed earlier. These operators have scaling dimensions which can be inferred from radial scaling of the supergravity perturbations, or can also be protected and therefore determined directly in the CFT. We also state the origin of the perturbation in the supergravity, following the conventions laid out in \cite{Baumann:2010sx} and reviewed in the appendix. Finally, we express the quantum numbers of the operator under the CFT global symmetry group $SU(2) \times SU(2) \times U(1)_R$\footnote{Note that in KW and its orbifold, the R-symmetry is exact. However, instantons in KS break this symmetry, and therefore could presumably induce dangerous operators. We treat operators in the UV, KW limit, and so we should not expect to be able to generate operators which violate the R-symmetry. Even if we were to discuss operator perturbations to KS, though, for $m \geq 2$, a group larger than R-parity is preserved, and so instantons would not be able to generate any operators in the superpotential that would threaten to destabilize the hierarchy.}.

A dangerous GSRO would be a singlet relevant single-trace primary operator or a double-trace primary operator which is the square of a (non-singlet) single-trace primary operator\footnote{The double-trace primary operator will have an anomalous dimension, but it will be $O\left(\frac{1}{n}\right)$ and so we ignore the correction.} with $\Delta < 2$. We see that there are no such primary operators of either kind in the spectrum, and therefore we conclude that the complex cone over $\F_0$ exhibits weak accidental SUSY. Drawing on this conclusion, we expect that the deformed complex cone over $\F_0$ exhibits weak approximate accidental SUSY, and therefore adding any nonsupersymmetric perturbations (which respect the global symmetries) to the Lagrangian give rise to an IR theory which is approximately supersymmetric. Therefore, this model is a proof-of-principle of the existence of a UV-complete prototype of a quasi-SCFT sector suitable for BSM model-building.


\section{Conclusions}
\label{sec:conclusions}

In this work, we have discussed the mechanism of accidental SUSY, describing QFTs which are nonsupersymmetric, but have supersymmetry appearing ``by accident'' in the IR. Accidental SUSY can be exact or approximate, depending on whether we reach a supersymmetric fixed point or simply flow towards one for a while, and can be strong or weak, depending on whether there are not or are global singlet marginal primary operators in the spectrum, respectively.

We discussed the model of \cite{Sundrum:2009gv}, which we refer to as resurgent SUSY. This model utilizes a weak approximate accidentally supersymmetric sector in a BSM model in order to partially UV-complete natural SUSY. We discuss that sector in detail and impose constraints on how such a sector must be realized in a full UV-completion.

We then move on to discuss what one must do in a string construction to realize even a prototype of the aforementioned accidentally supersymmetric sector in a fashion compatible with resurgent SUSY. From there, we move to study a SUSY-preserving $\Z_2$ orbifold of the noncompact Klebanov-Strassler theory. We were able to describe the space of global symmetry-respecting UV Lagrangian deformations of the dual gauge theory by classifying classical perturbations to the KW theory, and used the AdS/CFT correspondence to match those perturbations to primary operators in the gauge theory. We were therefore able to conclude that this orbifold theory admits weak approximate accidental supersymmetry, and is therefore a suitable prototype for the quasi-conformal sector in resurgent SUSY.

The theory discussed in this paper is over a noncompact Calabi-Yau, meaning that the theory does not have dynamical four-dimensional gravity. This is not a problem from the point of view of attempting to exhibit any UV-complete model of accidental SUSY, but one might hope to incorporate 4d gravity in more realistic string models which reproduced the MSSM or the natural SUSY spectrum. Due to a no-go theorem, a compactification to four dimensions is best described in the language of F-theory. In that language, the noncompact model described in this paper is an effective ``linearization'' of the warped throat present in the full compactification. By allowing for non-normalizable perturbations in supergravity, we systematically allow any and all operators respecting the $\Z_2$ and continuous global symmetries described in \ref{sec:gauge}, knowing that all such operators cannot effectively transmit SUSY-breaking.

In future work, we plan to pursue the question of constructing an F-theory model which reproduces the model in this paper in some local patch of the base of the Calabi-Yau fourfold. It is not immediately obvious that such a fourfold should exist, but the existence of F-theory models such as described in \cite{Giddings:2001yu, Kachru:2009kg} gives us hope that it is feasible. Even if this particular noncompact construction cannot be realized in F-theory, it is nevertheless interesting to ask about the possibility of other F-theory models which do exhibit accidental supersymmetry. One would like the F-theory compactification which locally reproduces the model described in this paper to satisfy the following checklist:

\begin{itemize}
\item It should contain a warped throat which supports an adequately large hierarchy, meaning that the Euler number of the fourfold should be sufficiently large.
\item It must respect the two $\Z_2$ symmetries described in section \ref{sec:gauge}, which are necessary to protect against GSROs.
\item Compact Calabi-Yaus of adequately general holonomy do not possess continuous isometries, and so we must preserve a sufficiently large discrete subgroup of $SU(2)\times SU(2)\times U(1)_R$ to prevent those non-global singlet operators described in table \ref{tab:matching} from being generated.
\item We must break SUSY in the bulk of the Calabi-Yau.
\end{itemize}

There are, of course, the usual worries about moduli stabilization; we assume that the F-theory fluxes stabilizes the moduli of the compactification. It would be interesting to explore general string corrections to this model; in particular, it would be interesting to explore the breaking of the no-scale structure present in the supergravity.

Furthermore, it would be interesting to attempt to modify the model presented here in order to make it more realistic; one would like to add weak gauge groups by stretching D7s down the length of the throat, and add a composite Higgs which breaks some of the gauge groups. These are necessary because the F-theory compactification will prevent us from utilizing the continuous isometries in our noncompact case as proxies for the SM gauge group.

Finally, although the question of individual models of accidental SUSY are interesting in their own right, it would be very exciting to have a sense of how ``generic'' features such as accidental SUSY are on the landscape, defined in relation to the number of vacua with high-scale SUSY-breaking and a finely-tuned electroweak scale. It was argued that warped throats are fairly ubiquitous in string theory \cite{Hebecker:2006bn}; it would be very interesting if accidental SUSY were as well. We hope that tests of the presence of warped throats or accidental SUSY can be developed and utilized to scan the landscape, in a similar spirit to the surveys of \cite{Anderson:2011ns, Anderson:2012yf, Anderson:2013xka}, as inspiration to pursue the idea of naturalness further at the 14 TeV LHC and beyond, as well as open new chapters in the field of string phenomenology.

\appendix

\section{Conventions}

{\bf Notations and conventions}

We use the mostly-plus metric signature. $d^4x$ is shorthand for $dx^0 \wedge dx^1 \wedge dx^2 \wedge dx^3$. $\kappa$, $\alpha'$ and $g_s$ are related by $2\kappa^2 = (2\pi)^7 \alpha'^4 g_s^2$. We switch between setting the string length $l_s = 1$ ($\alpha' = \frac{1}{2}$) and setting the AdS radius of curvature $R=1$, but attempt to indicate when we have done which. $g_s = \langle e^\phi \rangle = \langle \frac{1}{\im\tau}\rangle$.\\

{\noindent \bf Gauge transformations}

In KW/KS, we use the following conventions for superfield gauge transformations:

\begin{equation}e^{V_1} \rightarrow e^{-i\bar{\Lambda}_1}e^{V_1} e^{i\Lambda_1}\qquad e^{V_2} \rightarrow e^{-i\bar{\Lambda}_2}e^{V_2} e^{i\Lambda_2}\end{equation}
\begin{equation}W_1 \rightarrow e^{-i\Lambda_1} W_1 e^{i\Lambda_1} \qquad W_2 \rightarrow e^{-i\Lambda_2} W_2 e^{i\Lambda_2} \end{equation}
\begin{equation}A \rightarrow e^{-i\Lambda_1} A e^{i\Lambda_2} \qquad B \rightarrow e^{-i\Lambda_2} B e^{i\Lambda_1}\end{equation}\\

{\noindent \bf IIB Supergravity}

We define the following combination of IIB supergravity fields:

\begin{equation}\tau = C_0 + i e^{-\phi} \qquad G_3 = F_3 - \tau H_3 \qquad \tilde{F}_5 = F_5 - \frac{1}{2} C_2 \wedge H_3 + \frac{1}{2} B_2 \wedge F_3 \end{equation}

The IIB Einstein-frame action is \cite{Bergshoeff:1995sq}:

\begin{equation} S =  \frac{1}{2\kappa^2} \int \Bigg(R*1 - \frac{d\tau \wedge *d\bar{\tau}}{2(\im \tau)^2} - \frac{G_3 \wedge *\bar{G}_3}{2\im\tau} - \frac{\tilde{F}_5 \wedge *\tilde{F}_5}{4}  - \frac{C_4 \wedge G_3 \wedge \bar{G}_3}{4i\im\tau} \Bigg)\end{equation}

\noindent where under a modular $SL(2,\Z)$ transformation,

\begin{equation}\tau \rightarrow \frac{a\tau + b}{c\tau + e}\qquad G_3 \rightarrow \frac{G_3}{c\tau + e}\end{equation}

\noindent and every other field is invariant. The associated equations of motion and Bianchi identities are \cite{Schwarz:1983qr}:

\begin{equation}-\frac{2}{\im\tau}d\tau \wedge *d\tau + 2id*d\tau = G_3 \wedge *G_3\end{equation}

\begin{equation} \frac{d\tau \wedge * (G_3+\bar{G}_3)}{2\im\tau}- i d*G_3 = \tilde{F}_5 \wedge G_3 \end{equation}

\begin{equation}d*\tilde{F}_5 = d\tilde{F}_5 = -\frac{G_3 \wedge \bar{G}_3}{2i\im\tau}\end{equation}

\begin{equation}dG_3 = -d\tau \wedge H_3\end{equation}

\begin{equation}R_{MN} = \frac{\partial_M \tau \partial_N \bar{\tau}}{2(\im\tau)^2} + \frac{G_{3MPQ}G_{3N}^{\phantom{3N}PQ}}{4\im\tau} + \frac{\tilde{F}_{5MPQRS}\tilde{F}_{5N}^{\phantom{5N}PQRS}}{96} -g_{MN}\frac{|G_3|^2}{48\im\tau}\end{equation}

Finally, we must impose by hand that $\tilde{F}_5 = *\tilde{F}_5$. Although the above equations of motion are consistent with self-duality, they do not imply it.

We often work with the metric and fiveform flux ans\"atze

\begin{equation}ds^2 = e^{2A}\eta_{\mu\nu}dx^\mu dx^\nu + e^{-2A}g_{mn}ds_{M_6}^2 = e^{2A}\eta_{\mu\nu}dx^\mu dx^\nu + e^{-2A}(dr^2 + r^2 ds_{X_5}^2)\end{equation}

\begin{equation}\tilde{F}_5 = (1+*)d\alpha \wedge d^4x\end{equation}

\noindent in terms of which we can define the fields $\Phi_\pm = e^{4A}\pm \alpha$. The equations of motion above become

\begin{equation}\Box_6 \Phi_- = \frac{e^{2A}}{6\im\tau}|2P_-G_3|^2 + e^{-4A}g^{mn}\partial_m\Phi_- \partial_n \Phi_-\end{equation}
\begin{equation}\Box_6 \Phi_+ = \frac{e^{2A}}{6\im\tau}|2P_+G_3|^2 + e^{-4A}g^{mn}\partial_m\Phi_+ \partial_n \Phi_+\end{equation}

\noindent where both $\Box_6$ and $\partial_m$ are to be evaluated with respect to the $M_6$ metric (without warp factors). In KW, the solutions are $\Phi_- = 0$, $\Phi_+ = 2r^4$. We have introduced the projection operators $P_{\pm}$ which project threeforms onto their imaginary self-dual and imaginary anti-self dual (ISD and IASD, respectively) components. These operators are defined by

\begin{equation}P_{\pm} = \frac{1}{2i}(i \pm *_6)\end{equation}

\noindent where $*_6$ only dualizes with respect to $M_6$ and $g_{mn}$. These operators satisfy

\begin{equation}P_+^2 = P_+\qquad P_-^2 = P_- \qquad P_+P_- = P_-P_+ = 0 \qquad P_+ + P_- = 1\end{equation}\\

{\noindent \bf The conifold}

The conifold is a noncompact Calabi-Yau threefold, and it can be described as the subspace of $\mathbb{C}^4$ satisfying

\begin{equation}z_1^2 + z_2^2 + z_3^2 + z_4^2 = 0\end{equation}

Fixing the radial distance from the origin, the horizon manifold $X_5$ is the five-dimensional Sasaki-Einstein space $T^{1,1}= \frac{S^3\times S^3}{S^1_{diag}}$. $T^{1,1}$ can be described by coordinates $(\psi, \theta_1, \theta_2, \phi_1, \phi_2)$ \cite{Candelas:1989js}, and is topologically $S^2 \times S^3$. Here, $\theta_i$ runs from $0$ to $\pi$, $\phi_i$ from $0$ to $2\pi$, and $\psi$ from $0$ to $4\pi$. In terms of these coordinates, the Sasaki-Einstein metric is

\begin{equation}ds^2 = \frac{1}{9} \left(d\psi + \cos\theta_1 d\phi_1 + \cos\theta_2 d\phi_2\right)^2 + \frac{1}{6}\left(d\theta_1^2 + d\theta_2^2 + \sin^2\theta_1 d\phi_1^2 + \sin^2\theta_2 d\phi_2^2\right)\end{equation}

The metric can be rewritten diagonally in terms of the globally-defined one-forms $g_{1...5}$, related to the coordinates above by

\begin{align}g_1 &= \frac{1}{\sqrt{2}}\left(-\sin\theta_1 d\phi_1 - \cos\psi \sin\theta_2 d\phi_2 + \sin\psi d\theta_2\right)\\ \nonumber
g_2 &= \frac{1}{\sqrt{2}}\left(d\theta_1 - \sin\psi \sin\theta_2 d\phi_2 - \cos\psi d\theta_2\right)\\ \nonumber
g_3 &= \frac{1}{\sqrt{2}}\left(-\sin\theta_1 d\phi_1 + \cos\psi \sin\theta_2 d\phi_2 - \sin\psi d\theta_2\right)\\ \nonumber
g_4 &= \frac{1}{\sqrt{2}}\left(d\theta_1 + \sin\psi \sin\theta_2 d\phi_2 + \cos\psi d\theta_2\right)\\ \nonumber
g_5 &= d\psi + \cos\theta_1d\phi_1 + \cos\theta_2 d\phi_2\end{align}

The metric in these coordinates is

\begin{equation}ds^2 = \frac{1}{6}\left(g_1^2 + g_2^2 + g_3^2 + g_4^2\right) + \frac{1}{9}g_5^2\end{equation}

It is convenient to define the following:

\begin{equation}\omega_2 = \frac{1}{2}\left(\sin \theta_1 d\theta_1 \wedge d\phi_1 - \sin\theta_2 d\theta_2 \wedge d\phi_2\right)\end{equation}
\begin{equation}\omega_3 = g_5 \wedge \omega_2\end{equation}

We use $\omega_5$ to denote the volume form; it is

\begin{equation}\omega_5 = \frac{1}{54}\omega_2 \wedge \omega_3 = \frac{1}{108}\sin\theta_1 \sin\theta_2 d\psi\wedge d\theta_1 \wedge d\theta_2 \wedge d\phi_1 \wedge d\phi_2\end{equation}

\noindent Integrating this over $T^{1,1}$ gives the volume $V = \frac{16\pi^3}{27}$.

{\noindent \bf The Klebanov-Witten theory}

The near-horizon limit of the Klebanov-Witten solution to IIB supergravity is

\begin{equation}ds^2 = e^{2A} \eta_{\mu\nu} dx^\mu dx^\nu + e^{-2A} (dr^2 + r^2 ds_{T^{1,1}}^2)\end{equation}

\begin{equation}A = \ln\frac{r}{R} \qquad  R = \left(\frac{27\pi g_sN}{4}\right)^{1/4} \qquad \tau = \frac{i}{g_s} \qquad G_3 = 0\end{equation}

\begin{equation}\tilde{F}_5 = (1+*)d\alpha \wedge d^4x \qquad \alpha = e^{4A} = \frac{r^4}{R^4} \end{equation}

\noindent where $R$ is the AdS radius of curvature measured in units of the string length, $l_s=1$. Note that $T^{1,1}$ is Sasaki-Einstein, implying the existence of $4d$ $\N = 1$ SUSY. The dual SCFT is described by gauge groups $SU(N)\times SU(N)$  along with bifundamental and antibifundamental fields $A^i$ and $B^j$, respectively, each transforming as a doublet under their own $SU(2)$ flavor symmetry group. At a strongly-coupled conformal fixed point, the scaling dimension of the matter fields is $\Delta = \frac{3}{4}$, and the theory has a superpotential 

\begin{equation}W = \lambda \epsilon_{ik}\epsilon_{jl} \tr(A^iB^jA^kB^l)\end{equation}

\noindent At weak coupling, the K\"ahler potential is given by

\begin{equation}K = \tr \left(\bar{A}e^{V_1}A e^{-V_2} + \bar{B}e^{V_2} B e^{-V_1}\right)\end{equation}

\section{Spectroscopy of IIB Supergravity on the Complex Cone Over $\F_0$}

We discuss the supergravity perturbation theory in subsection \ref{sec:sugra}, then turn to a categorization of operators in the dual gauge theory in subsection \ref{sec:gauge}. These are matched in the main text in table \ref{tab:matching}.

\subsection{The Supergravity Side}
\label{sec:sugra}

We consider a supergravity perturbation theory similar to that described in \cite{Baumann:2010sx}, allowing all fields to be systematically expanded in a formally small parameter\footnote{From the point of view of the dual gauge theory, this is a natural expansion to make; powers of $\varepsilon$ can be treated as numbers of spurion insertions.} $\varepsilon$ about their background values; e.g.

\begin{equation}\tilde{F}_5 = \tilde{F}_5^{(0)} + \varepsilon \tilde{F}_5^{(1)} + \ldots\end{equation}

In this case we want to allow all supergravity fields to be expanded, but insist that the resulting modes be $AdS_5$-scalar perturbations. Consequently, we ignore all fermion equations of motion. Furthermore, we would like the fluxes associated with the various gauge fields to not break 4d Poincar\'e-invariance. With respect to the near-brane KW background solution, the relevant\footnote{IIB supergravity also contains $\Phi_+$, but the perturbations of $\Phi_+$ dual to operators of the smallest dimension start at scaling dimension $\Delta = 6$, and are therefore irrelevant for our purposes \cite{Ceresole:1999zs, Baumann:2010sx}.} linearized perturbation equations become

\begin{equation}\Box_{(0)} \tau^{(1)} = 0 \qquad dG_3^{(1)}=0 \qquad d(r^4 P_- G_3^{(1)}) = 0 \qquad d\tilde{F}_5^{(1)}=0 \qquad \Box_{6(0)}\Phi_-^{(1)} = 0\end{equation}

In addition, there is the Einstein equation describing symmetric tensor perturbations of $T^{1,1}$; however, having found the spectrum of scalar operators in the CFT, we can deduce that any protected CFT operator not matched to one of the above supergravity fields must be dual to a symmetric tensor perturbation, and therefore do not need to solve the Einstein equation.

The process of solving the perturbation equations is greatly simplified by knowing the radial scaling of the scalar harmonics on $T^{1,1}$, which are known to be \cite{Ceresole:1999zs}

\begin{equation}\delta = -2 + \sqrt{H(j,l,r) + 4}\end{equation}

\noindent where $(j,l,r)$ describe the representation of the harmonic with respect to the isometry group of $T^{1,1}$, which is $SU(2) \times SU(2) \times U(1)$. $H$ is defined as

\begin{equation}H(j,l,r) = 6\left(j(j+1) + l(l+1) - \frac{1}{8}r^2\right)\end{equation}

\noindent The lowest values of $\delta$ are shown in table \ref{tab:scalarharmonics}. Note that for every nonvanishing value of $r$, there is another representation $(j,l,-r)$ with the same scaling dimension.

{\renewcommand{\arraystretch}{1.1}
\begin{table}[h] \centering \small
\begin{tabular}{|c|c|c|c|}
\hline
$\delta$ & $j$ & $l$ & $r$ \\ \hline \hline
0 & 0 & 0 & 0 \\ \hline
1.5 & $\frac{1}{2}$ & $\frac{1}{2}$ & 1 \\ \hline
2 & 1 & 0 & 0 \\ \hline
2 & 0 & 1 & 0 \\ \hline
3 & 1 & 1 & 2 \\ \hline
3.29 & 1 & 1 & 0 \\ \hline
3.5 & $\frac{3}{2}$ & $\frac{1}{2}$ & 1 \\ \hline
3.5 & $\frac{1}{2}$ & $\frac{3}{2}$ & 1 \\ \hline
\end{tabular}
\caption{The scaling dimensions and quantum numbers of the first few scalar harmonics on $T^{1,1}$.}
\label{tab:scalarharmonics}
\end{table}
}

The $\tau^{(1)}$ perturbation equation is independent of the rest and is satisfied by any harmonic scalar on $T^{1,1}$. We use coordinates $x$ for $AdS_5$ and $y$ for $T^{1,1}$; then we expand $\tau^{(1)}(x,y) =  \sum_{jlr} \phi_{jlr}(x) Y^{jlr}(y)$. The 10d Laplacian becomes $\Box_{AdS} - \Box_{T^{1,1}}$, where $\Box_{T^{1,1}} Y^{jlr}(y) = H(j,l,r)Y^{jlr}(y)$. Therefore, the AdS equation of motion for each term $\tau_{jlr}$ in the expansion becomes

\begin{equation}\Box_{AdS} \tau_{jlr}(x,y) = H(j,l,r) \tau_{jlr}(x,y)\end{equation}

\noindent These have mass-squareds beginning at $m^2 = H(0,0,0) = 0$ in $AdS_5$, and so with the exception of the constant mode, the rest cannot transmit SUSY-breaking from the UV as they're dual to operators with $\Delta > 4$. The constant mode is dual to an operator with $\Delta = 0$ or $4$, where these two choices are related by a Legendre transform \cite{Klebanov:1999tb}; the $\Delta = 0$ mode is a modulus which is the sum of the inverse-squared gauge couplings $1/g_1^2 + 1/g_2^2$ \cite{Klebanov:1998hh}.

{\renewcommand{\arraystretch}{1.1}
\begin{table}[h] \centering \small
\begin{tabular}{|c|c|c|c|}
\hline
$\Delta$ & $j$ & $l$ & $r$ \\ \hline \hline
4 & 0 & 0 & 0 \\ \hline
\end{tabular}
\caption{The scaling dimensions and quantum numbers of scalar harmonics of $\tau$ with $\Delta \leq 4$.}
\label{tab:tau}
\end{table}
}

By the Bianchi identity, $G_3^{(1)}$ is a closed threeform; therefore it is either exact or a representative of a cohomology class. We consider these in turn. As $\tau^{(0)}$ is constant, we can write $G_3$ as $dA_2$, where $A_2$ is the complex twoform $C_2 - \frac{i}{g_s} B_2$. Perturbations of $A_2$ must be $AdS_5$-scalars; therefore $A_2$ can be decomposed in terms of twoform harmonics on $T^{1,1}$. We write this decomposition as

\begin{equation}G_3^{(1)}(x,y) =  \sum_{i}d(f_i(x) \Omega_2^i(y))\end{equation}

\noindent where $\Omega_2^i$ are all harmonic two-forms on $T^{1,1}$, satisfying $*_5d\Omega_2 = i\delta \Omega_2$, where $*_5$ is the Hodge star operator with respect to $T^{1,1}$. $\delta$ are ($-i$ times) the eigenvalues of the Laplace-Beltrami operator. However, $f_i$ must be only a function of $r$ and not of 4d Minkowski coordinates; otherwise we would have a nontrivial flux along $\R^{1,3}$ in the CFT, breaking the Poincar\'e invariance of the CFT vacuum. We use the ansatz $f_i(x) = r^{\alpha_i}$, allowing for potentially multiple values of $\alpha$ for a given $i$. As we can solve the perturbation equation term-by-term, we must solve

\begin{equation}d(r^4 P_- (r^\alpha \Omega_2)) = 0 \end{equation}

The solution to this equation was found in \cite{Baumann:2010sx}; the claim is that our ansatz solves the perturbation equations when $\alpha = \delta -4$ or $-\delta$ (for $\delta \neq 0)$. In these cases, the scaling dimension of the dual operator\footnote{This rule is used to enforce unitarity of the dual operator ($\Delta \geq 1$). In the cases where both $4-\delta$ and $\delta$ are allowed by unitarity, either scaling dimension is acceptable in the dual CFT, the two choices being related by a Legendre transform \cite{Klebanov:1999tb}. We choose the larger scaling dimension for the duals of $G_3$-flux for ease in matching to the CFT spectrum.} is $\Delta = \mathrm{max}(4-\delta,\delta)$. The eigenvalues of the twoform harmonics were worked out in \cite{Ceresole:1999zs}; the answers can again be expressed in terms of the quantum numbers of the isometry group $(j,l,r)$. There are six eigenvalues of the Laplace-Beltrami operator for each $(j,l,r)$, and they are\footnote{Note that we disagree with the authors of \cite{Ceresole:1999zs} with regard to a minus sign.}

\begin{equation}\delta_{(j,l,r+2)} = 1 \pm \sqrt{H(j,l,r)+4}\end{equation}
\begin{equation}\delta_{(j,l,r-2)} = -1 \pm \sqrt{H(j,l,r)+4}\end{equation}
\begin{equation}\delta_{(j,l,r)} = \pm \sqrt{H(j,l,r)+4}\end{equation}

\noindent where the allowed values of $(j,l,r)$ and the definition of $H(j,l,r)$ are the same as in table \ref{tab:scalarharmonics}, but now the physical value of $r$ of the perturbation may not be $r$, but rather $r\pm 2$, as indicated in the subscript of $\delta$. This occurs because the twoform harmonics can be built out of scalar harmonics, but some of the solutions depend on the holomorphic threeform $\Omega$ on the conifold, which carries an $R$-charge of $2$.

The de Rham cohomology representative twoform $\omega_2$ of $T^{1,1}$ is an acceptable harmonic perturbation of the twoform gauge fields, corresponding to vanishing $G_3$. $\omega_2$ is therefore another modulus of the dual CFT; in this case it is the difference in the inverse gauge-squared couplings $\frac{1}{g_1^2}-\frac{1}{g_2^2}$. Again, this is the $\Delta = 0$ choice related to the $\Delta = 4$ choice by a Legendre transform.

The second singular cohomology of $T^{1,1}/\Z_2$ contains a $\Z_2$ torsion subgroup, related by Poincar\'e-duality to the presence of torsion threecycle in homology. However, an associated twoform is not present in the de Rham cohomology; $H^2_{\mathrm{dR}}(T^{1,1}/\Z_2,\R) = \R$. Consequently, there is no associated $A_2$-perturbation.

Finally, one can ask about $G_3$-flux in $H^3(T^{1,1})$. However, the cohomology representative $\omega_3 = g_5 \wedge \omega_2$ of $T^{1,1}$ does not satisfy $d(r^4 P_- \omega_3)=0$, and so it does not constitute an allowable perturbation of the solution.

{\renewcommand{\arraystretch}{1.1}
\begin{table}[h] \centering \small
\begin{tabular}{|c|c|c|c|c|}
\hline
Series & $\Delta$ & $j$ & $l$ & $r$ \\ \hline \hline
I & 2.5 & $\frac{1}{2}$ & $\frac{1}{2}$ & -1 \\ \hline
I & 3 & 0 & 0 & -2 \\ \hline
I & 3 & 1 & 0 & -2 \\ \hline
I & 3 & 0 & 1 & -2 \\ \hline
I & 4 & 1 & 1 & 0 \\ \hline
II & 2 & 0 & 0 & 0 \\ \hline
II & 3.5 & $\frac{1}{2}$ & $\frac{1}{2}$ & 1 \\ \hline
II & 4 & 1 & 0 & 0 \\ \hline
II & 4 & 0 & 1 & 0 \\ \hline
III & 3 & 0 & 0 & 2 \\ \hline
$\omega_2$ & 4 & 0 & 0 & 0 \\ \hline
\end{tabular}
\caption{The scaling dimensions and quantum numbers of scalar harmonics of $A_2$ with $\Delta \leq 4$.}
\label{tab:tau}
\end{table}
}

We turn to $\Phi_-$, satisfying $\Box_6 \Phi^{(1)}_- = 0$. We expand again as $\Phi^{(1)}_-= \sum_i r^{\alpha_i}Y_i$, and solve term by term. The solutions are

\begin{equation}\alpha = -2 \pm \sqrt{4+H}\end{equation}

\noindent where $\Box_5 Y = H Y$. The dual operators' scaling dimensions are therefore $\Delta = -2 + \sqrt{4+H}$, and we list those modes in table \ref{tab:phiminus}.

{\renewcommand{\arraystretch}{1.1}
\begin{table}[h] \centering \small
\begin{tabular}{|c|c|c|c|}
\hline
$\Delta$ & $j$ & $l$ & $r$ \\ \hline \hline
1.5 & $\frac{1}{2}$ & $\frac{1}{2}$ & 1 \\ \hline
2 & 1 & 0 & 0 \\ \hline
2 & 0 & 1 & 0 \\ \hline
3 & 1 & 1 & 2 \\ \hline
3.29 & 1 & 1 & 0 \\ \hline
3.5 & $\frac{3}{2}$ & $\frac{1}{2}$ & 1 \\ \hline
3.5 & $\frac{1}{2}$ & $\frac{3}{2}$ & 1 \\ \hline
\end{tabular}
\caption{The scaling dimensions and quantum numbers of harmonics of $\Phi_-$ with $\Delta \leq 4$.}
\label{tab:phiminus}
\end{table}
}

Perturbations of the fiveform flux which are in the cohomology of $T^{1,1}$ clearly satisfy the equations of motion as $d\omega_5 = 0$ and $d*\omega_5 = 0$; however, perturbing the fiveform flux in KW simply changes the number of D3-branes one has in the solution, dual to changing the rank of the gauge group.

\subsection{The Gauge Side}
\label{sec:gauge}

In this section, we study the gauge dual of the supersymmetry-preserving $\Z_2$-orbifold of Klebanov-Strassler. We denote $n = \frac{1}{2}N$ and $m = \frac{1}{2} M$, implicitly assuming that $N$ and $M$ are even. Recall that the orbifold action in the supergravity was $z_i \rightarrow -z_i$, where $z \sim AB$. $A$ and $B$ are bifundamentals under the gauge group, and can therefore be written as $N\times(N+M)$ and $(N+M)\times N$ matrices, respectively. The action of the orbifold can therefore be embedded on the gauge side as follows:

\begin{itemize}
{\item $SU(N+M)$ is orbifolded by diag$(I_{n+m},-I_{n+m})$. The resulting groups are two copies of $SU(n+m)$, and these are called groups $G_1$ and $G_3$, respectively, as well as a $U(1)$.}
{\item $SU(N)$ is orbifolded by diag$(I_{n},-I_{n})$. The resulting groups are two copies of $SU(n)$, and these are called groups $G_2$ and $G_4$, respectively, as well as another $U(1)$.}
{\item The gauginos are embedded the same way as the gauge bosons so as to preserve supersymmetry.}
{\item The two superfields $A^i$ are odd under the orbifold. The resulting superfields are embedded in $A^i$ as:
\begin{equation}\left(\begin{matrix}0 & Q_1^i \\ Q_3^i & 0\end{matrix}\right)\end{equation} }
{\item The two superfields $B^j$ are even under the orbifold. The resulting superfields are embedded in $B^j$ as:
\begin{equation}\left(\begin{matrix}Q_2^i & 0 \\ 0 & Q_4^i\end{matrix}\right)\end{equation} }
\end{itemize}

Under the action of the orbifold, the superpotential can be written as

\begin{equation}W = \lambda \varepsilon_{ik}\varepsilon_{jl}~\tr\left(Q_1^i Q_2^l Q_3^k Q_4^j\right)\end{equation}

where $\varepsilon_{12}=1$. The representations under the various groups are listed in table \ref{tab:orbifoldreps}.

{\renewcommand{\arraystretch}{1.1}
\begin{table}[h] \centering \small
\begin{tabular}{|c|c|c|c|c|c|c|c|c|c|c|c|}\hline
Field & $G_1$ & $G_2$ & $G_3$ & $G_4$ & $SU(2)_1$ & $SU(2)_2$ & $B$ & $B_A$ & $B_B$ & $A$ & $R$ \\ \hline \hline
$Q_1$ & $\square$ & 1 & 1 & $\bar{\square}$ & 2 & 1 & 1 & 1 & 0 & 1 & $\frac{1}{2}$ \\ \hline
$Q_2$ & $\bar{\square}$ & $\square$ & 1 & 1 & 1 & 2 & -1 & 0 & 1 & 1 & $\frac{1}{2}$ \\ \hline
$Q_3$ & 1 & $\bar{\square}$ & $\square$ & 1 & 2 & 1 & 1 & -1 & 0 & 1 & $\frac{1}{2}$ \\ \hline
$Q_4$ & 1 & 1 & $\bar{\square}$ & $\square$  & 1 & 2 & -1 & 0 & -1 & 1 & $\frac{1}{2}$ \\ \hline
$\lambda$ & 1 & 1 & 1 & 1 & 1 & 1 & 0 & 0 & 0 & -4 & 0 \\ \hline
$\Lambda_{1,3}^{b_{1,3}}$ & 1 & 1 & 1 & 1 & 1 & 1 & 0 & 0 & 0 & $4n$ & $2m$ \\ \hline
$\Lambda_{2,4}^{b_{2,4}}$ & 1 & 1 & 1 & 1 & 1 & 1 & 0 & 0 & 0 & $4(n+m)$ & $-2m$ \\ \hline
\end{tabular}
\caption{The representations of the fields in the orbifold of KS. $G_1$ and $G_3$ are $SU(n+m)$, $G_2$ and $G_4$ are $SU(n)$, and $B$, $B_A$, $B_B$, $A$ and $R$ are all $U(1)$s.}
\label{tab:orbifoldreps}
\end{table}
}

{\renewcommand{\arraystretch}{1.1}
\begin{table}[h] \centering \small
\begin{tabular}{|c|c|c|c|c|c|c|c|c|c|c|c|}\hline
Field & $G_1$ & $G_2$ & $SU(2)_1$ & $SU(2)_2$ & $B$ & $A$ & $R$ \\ \hline \hline
$A$ & $\square$ & $\bar{\square}$ & 2 & 1 & 1 & 1 & $\frac{1}{2}$ \\ \hline
$B$ & $\bar{\square}$ & $\square$ & 1 & 2 & -1 & 1 & $\frac{1}{2}$ \\ \hline
$\lambda$ & 1 & 1 & 1 & 1 & 0 & -4 & 0 \\ \hline
\end{tabular}
\caption{The representations of the fields in the orbifold of KS. $G_1$ and $G_3$ are $SU(n+m)$, $G_2$ and $G_4$ are $SU(n)$, and $B$, $B_A$, $B_B$, $A$ and $R$ are all $U(1)$s.}
\label{tab:orbifoldreps2}
\end{table}
}

The $U(1)_A$ is a spurious symmetry\footnote{This statement is only true perturbatively; instantons would break $U(1)_A$ even in the absence of a superpotential.}, with the superpotential coupling $\lambda$ acting as the spurion. Furthermore, the $U(1)_R$ symmetry is anomalous; the exact symmetry is $\Z_{2m}$. We also record the charges of the holomorphic intrinsic scales $\Lambda_i$, which we introduce shortly.

There are also two additional gauged baryon numbers $B_A$ and $B_B$ which appear in the orbifolded theory; these are linear combinations of the two daughter $U(1)$s of the parent gauge groups. These baryon numbers rotate the daughters of $A$ and the daughters of $B$ into each other, respectively. However, these groups both have mixed anomalies with the nonabelian gauge groups. This is a common occurrence in orbifold gauge theories \cite{vonGersdorff:2003dt}, and these anomalies do not pose a problem to us. However, as discussed in section \ref{sec:accsusy}, global $U(1)$s have conserved currents with scalar superpartners which generically are GSROs, and so it is instructive to see why the associated global symmetry currents do not cause trouble.

What happens here is that a generalized Green-Schwarz mechanism ``cancels'' the anomalies by restoring the gauge-invariance of the effective action and giving the $U(1)$ gauge bosons Stueckelberg masses. However, there remain two global $U(1)$s with scalar superpartners. Nevertheless, despite the theory being unitary and gauge-invariant, these $U(1)$s remain anomalous because the the divergence of the associated currents are nonvanishing. As described in \cite{Preskill:1990fr}, this is self-consistent as an effective theory, though this description requires a UV completion. In our case, the UV completion is given by the string description.

This is consistent with the dual string picture; in the orbifolded theory, a new sector of twisted states arise. However, because the orbifold is freely acting, there are no new massless states in the twisted sector. Therefore, we conclude that there are no new gauge bosons in the $AdS_5$ description and no new conserved currents in the daughter gauge theory description.

There is a residual $\Z_2$ symmetry of the quiver that exchanges groups $1$ and $3$ and also groups $2$ and $4$, along with an associated swapping of matter fields. We opt to impose that symmetry on all of our perturbations as deviations will lead to different physics in IR as compared to KS. This will place a restriction on the allowed sorts of supersymmetry-breaking in the compactification. Crucially, this forces the equality of gauge couplings $g_1$ and $g_3$, as well as $g_2$ and $g_4$.

Furthermore, there is a $\Z_2$ outer automorphism of KW that exchanges the two gauge groups as well as $A$ and $B$. The orbifold inherits this automorphism when $m=0$; in fact, it combines with the $\Z_2$ of the previous paragraph to form a full $D_4$ symmetry group \cite{Morrison:1998cs}, though we will not need the full $D_4$ for the purposes of this paper. The $\Z_2$ needs to be respected by SUSY-breaking in the compactification, as this prevents us from deforming the Lagrangian with the $U(1)_B$ scalar current with protected scaling dimension $\Delta = 2$. 

The NSVZ exact beta function for SQCD with $N$ colors and $F$ vector-like flavors is \cite{Novikov:1983uc}:

\begin{equation}\beta_g \equiv \mu \frac{dg}{d\mu} = -\frac{g^3}{16\pi^2} \frac{\left(3N-F(1-\gamma)\right)}{1-\frac{g^2N}{8\pi^2}} \equiv \frac{b^e g^3}{16\pi^2}\frac{1}{1-\frac{g^2N}{8\pi^2}}\end{equation}

\noindent where $\gamma$ is the anomalous dimension\footnote{Here we mean the conventional definition of ``anomalous dimension'', as opposed to the deviation due to $\frac{1}{N}$-corrections as we used in section \ref{sec:accsusy}.} of a quark field, related to the scaling dimension $\Delta$ of a field with engineering dimension $d$ by $\Delta = d + \frac{1}{2}\gamma$. As with KS, there is a UV fixed point when $m=0$. There, the anomalous dimension of the quark superfields are $\gamma = -\frac{1}{2}$ \cite{Morrison:1998cs, Franco:2003ja}, and there are two marginal couplings corresponding to a two-parameter family of fixed points. Here, we choose to sit near the fixed point, so $\gamma_i = -\frac{1}{2} + O\left(\frac{m^2}{n^2}\right)$. The vanishing of the first-order term arises because of a symmetry $m\rightarrow -m$, $n\rightarrow n+m$ of the theory. For groups $1$ and $3$, we have $n+m$ colors, and we effectively have $2n$ vector-like flavors. For groups 2 and 4, we have $n$ colors and $2(n+m)$ vector-like flavors. To leading order in $\frac{m}{n}$, then, we have $b^e_1 = b^e_3 = -3m$ and $b^e_2 = b^e_4 = 3m$. Thus, gauge groups 1 and 3 are UV-free, meaning that as we flow into the IR, they confine at some scale $|\Lambda_1|$, $|\Lambda_3|$ respectively. Groups 2 and 4, on the other hand, are IR-free, so the gauge couplings flow towards zero as we flow into the IR. As with SQCD, we introduce the holomorphic confinement scales

\begin{equation}\Lambda_i = |\Lambda_i| e^{\frac{i\theta_i}{b_i}}\end{equation}

\noindent where $\theta_i$ is the theta-angle of the $i$-th group and $b_i=-3N+F$ is the one-loop beta function coefficient. In our model, we have $b_1=b_3=n+3m$ and $b_2=b_4=n-2m$.

If we impose for our UV-free gauge groups $g_{1,3} = g_0$ at $\mu = \mu_0$, we reach a Landau pole as we flow into the IR at $\mu = |\Lambda|$, where

\begin{equation}|\Lambda| = \mu_0 e^{-\frac{8\pi^2}{3mg_0^2}}\end{equation}

The parent KS theory has an RG cascade, where one performs a Seiberg dual on the confining gauge group and continue flowing into the IR \cite{Seiberg:1994pq}. The orbifolded theory we've constructed inherits this RG cascade when $g_1=g_3$ and $g_2=g_4$. The dual theory is identical to the original theory, with $n+m$ replaced with $n-m$, the $U(1)_B$ and $U(1)_A$ charges rescaled, and the coupling $\lambda$ inverted. The case where the gauge couplings do not start out equal results in a dual theory which is not self-similar \cite{Franco:2003ja, Franco:2004jz, Franco:2005fd}, but this is not relevant for our purposes.

{\renewcommand{\arraystretch}{1.1}
\begin{table}[h] \centering \small
\begin{tabular}{|c|c|c|c|c|c|c|c|c|c|c|c|}\hline
Field & $G'_1$ & $G_2$ & $G'_3$ & $G_4$ & $SU(2)_1$ & $SU(2)_2$ & $B$ & $B_A$ & $B_B$ & $A$ & $R$ \\ \hline \hline
$Q_1'$ & $\bar{\square}$ & 1 & 1 & $\square$ & 2 & 1 & $\frac{n+m}{n-m}$ & $\frac{n+m}{n-m}$ & 0 & $\frac{n+m}{n-m}$ & $\frac{1}{2}$ \\ \hline
$Q_2'$ & $\square$ & $\bar{\square}$ & 1 & 1 & 1 & 2 & $-\frac{n+m}{n-m}$ & 0 & $\frac{n+m}{n-m}$ & $\frac{n+m}{n-m}$ & $\frac{1}{2}$ \\ \hline
$Q_3'$ & 1 & $\square$ & $\bar{\square}$ & 1 & 2 & 1 & $\frac{n+m}{n-m}$ & -$\frac{n+m}{n-m}$ & 0 & $\frac{n+m}{n-m}$ & $\frac{1}{2}$ \\ \hline
$Q_4'$ & 1 & 1 & $\square$ & $\bar{\square}$  & 1 & 2 & $-\frac{n+m}{n-m}$ & 0 & -$\frac{n+m}{n-m}$ & $\frac{n+m}{n-m}$ & $\frac{1}{2}$ \\ \hline
$M_{12}'$ & 1 & $\square$ & 1 & $\bar{\square}$ & 2 & 2 & 0 & 0 & 0 & 2 & 1 \\ \hline
$M_{34}'$ & 1 & $\bar{\square}$ & 1 & $\square$ & 2 & 2 & 0 & 0 & 0 & 2 & 1 \\ \hline
$\lambda'$ & 1 & 1 & 1 & 1 & 1 & 1 & 0 & 0 & 0 & $-4\frac{n+m}{n-m}$ & 0 \\ \hline
$\Lambda'_{1,3} \phantom{}^{b'_{1,3}}$ & 1 & 1 & 1 & 1 & 1 & 1 & 0 & 0 & 0 & $4n\frac{n+m}{n-m}$ & $-2m$ \\ \hline
$\Lambda_{2,4}^{b'_{2,4}}$ & 1 & 1 & 1 & 1 & 1 & 1 & 0 & 0 & 0 & $4(n+m)$ & $2m$ \\ \hline
\end{tabular}
\caption{The representations of the fields in the orbifold of KS after a Seiberg dual. $G'_1$ and $G'_3$ are $SU(n-m)$, $G_2$ and $G_4$ are $SU(n)$, and $B$, $B_A$, $B_B$, $A$ and $R$ are all $U(1)$s.}
\label{tab:dualorbifoldreps}
\end{table}
}

We list the field content of the dual in table \ref{tab:dualorbifoldreps}, where $G'_{1,3}$ are $SU(n-m)$ and $B$, $A$, etc. are again $U(1)$s. The charges under the $U(1)$s were determined by anomaly matching. The $SU(2)_1$, $SU(2)_2$ and $U(1)_B$ anomalies should match exactly. From the point of view of groups 1 and 3, groups 2 and 4 are flavor symmetries, and therefore we should match those anomalies as well. However, the $U(1)_{B_A}$, $U(1)_{B_B}$, $U(1)_A$ and $U(1)_R$ symmetries are anomalous\footnote{Of course, $U(1)_A$ isn't even a symmetry in the first place; regardless, it's a useful check to match anomaly coefficients.}; therefore, we only require the anomaly coefficients to match up to actual symmetry transformations\footnote{In order to utilize the results of \cite{Csaki:1997aw}, one needs to rescale $U(1)$ charges to be all integers.} \cite{Csaki:1997aw}. The $U(1)_R$ is broken to $\Z_{2m}$, and $U(1)_A$ is broken to the intersection of $\Z_{4(n+m)}$ and $\Z_{4n}$. That intersection is $\Z_4$ when $n+m$ and $n$ are coprime, but is $\Z_{4k}$ for some natural number $k$ when $n+m$ and $n$ are not coprime. Finally, note that we don't need to match anomalies with gauge groups 1 or 3, because the anomaly matching trick would require the addition of spectators that were charged under those groups, which changes the details of the confinement. Up to the above caveats, all anomaly coefficients match, lending credence to the idea of a duality cascade.

In the dual theory, there is a superpotential inherited from the original theory, which is self-similar after integrating out the mesons:

\begin{equation}W =\lambda' \varepsilon_{ik}\varepsilon_{jl} \tr \left( Q'_1\phantom{}^i Q'_2\phantom{}^l Q'_3\phantom{}^k Q'_4\phantom{}^j\right) \end{equation}

Now, groups 2 and 4 are UV-free and groups 1 and 3 are IR-free, and so the cascade continues until we cannot dualize any more, at which point the story plays out in a similar fashion to KS, with the deformation of the complex cone and the introduction of an ADS superpotential.

Such a duality cascade, complete with chiral symmetry breaking in the IR, is a fantastically rich IR, with lots of opportunities for model-building.

As with supergravity, we can classify operators by setting $m=0$ and studying their quantum numbers in the high-energy theory. The allowed operators fall into representations of the superconformal symmetry algebra $su(2,2|1)$ plus the global symmetry algebra $su(2) + su(2) + u(1)_R$. Note also that we ignore the global baryon number $B$, as all combinations of the fields which are gauge-invariant are automatically $B$-singlets. Chiral primary operators formed from various gauge and matter fields can be determined at weak coupling\footnote{We are concerned with the spectrum at strong coupling; many of the operators we consider have protected scaling dimension and therefore their fixed-point scaling dimension can be determined. This approach misses possible mixing between various primaries and descendants, but these are irrelevant for the purposes of spectroscopy.} and are subject to the following constraints: 

\begin{itemize}
\item They must be gauge-invariant.
\item The $F$-term equations of motion kill various potential flavor-singlet operators:
\begin{equation}\epsilon_{ik}A^i B^j A^k = 0 \qquad \epsilon_{jl} B^j A^i B^l = 0\end{equation}
\item The $D$-term equations of motion are
\begin{align}0 &= \tr \left(T_1^a (A^i e^{-V_2} \bar{A}_i - \bar{B}^j e^{V_2} B_j) \right) \nonumber \\
0 &= \tr \left(T_2^b (\bar{A}^i e^{V_1} A_i - B^j e^{-V_1} \bar{B}_j) \right) \end{align}
telling us that the adjoint operators $A^i e^{-V_2} \bar{A}_i - \bar{B}^j e^{V_2} B_j$ and $\bar{A}^i e^{V_1} A_i - B^j e^{-V_1} \bar{B}_j$ vanish in the supersymmetric vacuum, and only the associated singlets can be used to build operators. However, other than these operators themselves, any other operator we could build from them will necessarily be double-trace, and therefore irrelevant to our discussion.
\item The super-equations of motion reduce the number of chiral primaries. They are
\begin{equation}\bar{D}^2(e^{-V_2}\bar{A} e^{V_1}) = 0 \qquad \bar{D}^2(e^{-V_1}\bar{B}e^{V_2}) = 0\end{equation}
\item One must consider that various ``commutator'' operators may not, in fact, be chiral primaries\footnote{A more accurate way to phrase this would be to say that these operators are $\bar{D}$-cohomologous to zero, or that they vanish in the chiral ring.} due to superspace identities such as
\begin{equation}W_{1\alpha} A B - A W_{2\alpha}B = -\frac{1}{4} \bar{D}^2\left( e^{-V_1} D_\alpha \left( e^{V_1} A e^{-V_2}\right) e^{V_2} B\right)\end{equation}
\item The one-$\theta$ components of the $W$ are real; therefore they equal the one-$\bar{\theta}$ components of the $\bar{W}$:
\begin{equation}D^\alpha W_\alpha = \bar{D}_{\dot{\alpha}} \bar{W}^{\dot{\alpha}}\end{equation}
\end{itemize}

Superfields have protected scaling dimensions if they fall into one of the following categories:

\begin{itemize}
\item They are chiral; $\bar{D}_{\dot{\alpha}}X = 0$.
\item They are semichiral; $\bar{D}_{(\dot{\alpha}} X_{\dot{\beta_1}\ldots \dot{\beta_n})\beta_1\ldots} = 0$.
\item They are conserved; $D^\alpha X_{\alpha\ldots\dot{\alpha}\ldots} = \bar{D}^{\dot{\alpha}}X_{\alpha\ldots\dot{\alpha}\ldots} = 0$ or $\bar{D}^2 X_{\alpha\ldots} = 0$.
\item They are semiconserved; $\bar{D}^{\dot{\alpha}}X_{\alpha\ldots\dot{\alpha}\ldots} = 0$.
\end{itemize}

The classification of the operators in KW was carried out in \cite{Klebanov:1998hh, Ceresole:1999zs, Ceresole:1999ht}; we list the results for the chiral primaries with scalar components which have scaling dimension $\Delta \leq 4$, using the convention $(j,l,r)$ for the global quantum numbers to match the supergravity solutions. Note that non-real operators have hermitian conjugates with $(j,l,-r)$ and the same $\Delta$ which are excluded from the below list for brevity. This list can be found in table \ref{tab:matching}, matched to their dual supergravity modes.

\acknowledgments{We are grateful to Nima Arkani-Hamed, Jaume Gomis, Shamit Kachru, David E. Kaplan, Jared Kaplan, Heeyeon Kim, Simon Knapen, Liam McAllister and Matthew T. Walters for helpful discussions. We also thank Cyrus Faroughy and Matthew T. Walters for helpful feedback on various drafts of this manuscript. We are especially grateful to Raman Sundrum for very helpful discussions, extensive feedback on this draft and collaboration on early stages of this project. CB was supported under NSF grant PHY-1315155, and is also grateful for support from the Maryland Center of Fundamental Physics. Finally, the research of CB was supported in part by
the Perimeter Institute for Theoretical Physics. Research at Perimeter Institute is supported by the Government of Canada through Industry Canada and by the Province of Ontario through the Ministry of Economic Development and Innovation.}

\bibliography{asst_reformatted_refs}
\bibliographystyle{JHEP}

\end{document}